\documentclass[11pt,a4paper,twocolumn]{article}

\usepackage{anysize}    %pour les marges plus grosses a droite et a gauche
\usepackage{
   graphicx
}
\usepackage[hmargin=1.5cm,vmargin=1.5cm,headsep=0.5cm]{geometry}
\usepackage{url} %for bibliography WEB references (pmx-2008-11-04)

\usepackage[normalem]{ulem} %for underscore and line through text:
\newcommand{\name}[1]{\ensuremath{\mathit{#1}}}            % multiletter variables
\usepackage{float}
\restylefloat{figure,figure*,table}

\title{Reconfiguration of a Distributed Information Fusion System}
\author{\'Eric Benoit, Marc-Philippe Huget,\\
Patrice Moreaux, Olivier Passalacqua\\
LISTIC, Polytech'Savoie,\\
Universit{\'e} de Savoie,\\
74944 Annecy le Vieux, France\\
(e-mail: firstname.lastname@univ-savoie.fr)}

\begin{document}

\maketitle

\begin{abstract}                % Abstract of not more than 250 words.
Information Fusion Systems are now widely used in different fusion contexts,
like scientific processing, sensor networks, video and image processing.
One of the current trends in this area is to cope with distributed systems.
In this context, we have defined and implemented
a Dynamic Distributed Information Fusion System runtime model.
It allows us to cope with dynamic execution supports while trying to maintain the
functionalities of a given Dynamic Distributed Information Fusion System.
The paper presents our system, the reconfiguration problems we are faced with and our
solutions.
\end{abstract}

\paragraph{keywords}
Availability,
Data fusion,
Decision making,
Discrete-event dynamic systems,
Performance evaluation,
Run-time systems.

%file: intro.tex
%encoding : utf8
%description: introduction,  DCDS09 reconfiguration
%date: 2009-04-16
%authors: Olivier PASSALACQUA
%email: olivier.passalacqua@univ-savoie.fr
%LISTIC, Polytech'Savoie-Annecy, Universit\'e de Savoie
%LaTeX2e

\section{Introduction}
The aim of an Information Fusion System is to
compute \emph{results of higher quality} (with respect to some criteria to be defined)
from information provided to it either from the "real world" (sensor networks)
or from computer sources (databases for instance).
Present IFS are frequently \emph{distributed} since data sources or/and
computation resources power are actually distributed.

Computer based Information Fusion Systems are widely diffused since a couple of decades
(see for instance [\cite{ARMaBl97,ARKaZhKa97}]).
Although there are already a lot of solutions to develop and to deploy
Distributed Information Fusion Systems (DIFS),
see~[\cite{TRHoMaMiJoVe07}] for a survey of solutions in the sensor network
area for instance,
most of them are restricted to specific application areas.
Moreover, they frequently assume that the execution support system is fixed.

The goal of our project is to define a runtime framework for
Dynamic Distributed Information Fusion System (DDIFS).
This framework is based on a model of the Fusion Process (FP)
and on a model of its derived run-time deployment.
These two models allow us to control the DDIFS:
\begin{itemize}
\item
the system restores a correct state after a run-time error;
\item
the system modifies itself
when it detects that a better configuration, in a sense to be detailed,
could be deployed.
\end{itemize}
These two adaptive behaviours explain why our systems are termed
Controlled Dynamic Distributed Information Fusion Systems (CDDIFS).

Fusion methods can be classified~[\cite{ARSa02}] into three groups, according to their domain:
probabilistic models such as Bayesian networks~[\cite{ARMaDu06}],
approximation methods which update a model of the environment
and take decisions based on the predicted next state
(for example Kalman filters~[\cite{ARSu06}]) and
interpolation  methods such as fuzzy logic approaches~[\cite{ARBlGeMa02}]
and neural networks~[\cite{ARPaCa03}].

In our context, an \emph{Information Fusion Process} (IFP) is defined
as a discrete data-flow graph the nodes of which are fusion functions
and the edges of which are links between function ports.
Ports of a fusion node are connected to input, output and parameters of the
fusion function.
A fusion function $f$ computes a tuple of output values
$Y = (y_1,...,y_K)$
from a tuple of input values $X=(x_1,...,x_I)$,
for a given vector $\theta = (\theta_1,...,\theta_J)$ of parameters:
$$
(y_1,...,y_K) = f_{(\theta_1,...,\theta_J)}(x_1,...,x_I).
$$
Distinction between input and parameter values comes from
the fact that an input value is used for only one computation of the outputs,
while a parameter is a sustained value, used for each computation, until it is modified.
Our model is termed discrete since the tuples are discrete data
and each function ``consumes'' one input tuple
and produces one output tuple.

Our approach terms \emph{Information Fusion System} (IFS)
 a hardware and software environment used to implement an IFP.
It includes all the components needed to execute the fusion function implementations
and to transfer information between the functions.
The distributed aspect comes from the physical distribution of the run-time elements
of the IFS: sensors, devices, computers, smart-phones, etc., are actually usually distributed.
We call \emph{execution framework} (EF) the computation environment where implementations
of fusion nodes run (see section~\ref{sec:implementation}).

\begin{figure*}
\begin{center}
\includegraphics[scale=1.2]{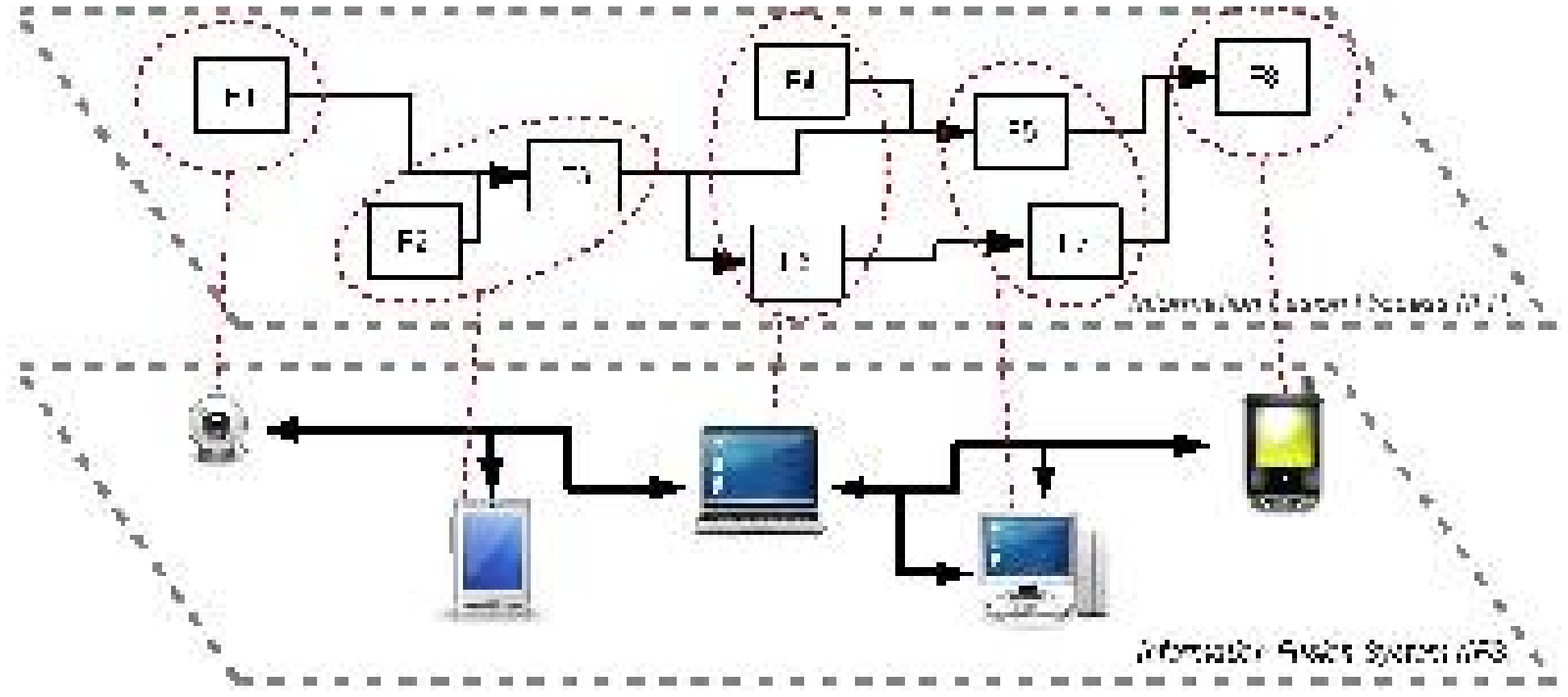}
\end{center}
\caption{Information Fusion Process (IFP) - Information Fusion System (IFS) relationship\label{fig:twolevelscontrol}}
\end{figure*}

Our model deals with dynamicity in the sense that it copes
with possible modifications of the IFS at run-time.
Modifications may be the update of a fusion node implementation,
the modification of network connections between the EFs
or the failure of a part of the IFS.
Note that however in this paper, it is assumed that the IFP is fixed.

The paper is organised as follows.
The next section explains how we control the IFS.
Section~\ref{sec:implementation} details our current implementation,
while in section~\ref{sec:reconfiguration}, we present the reconfiguration strategies
we have designed and their implementation.
An application example is detailed in section~\ref{sec:examples},
and we indicate work in progress in section~\ref{sec:conclu}.

% end of file intro.tex

%\input{context}
%file: control.tex
%description: control of distributed information fusion system, dcds09
%date: 2009-04-16
%author: Olivier PASSALACQUA
%email: olivier.passalacqua@univ-savoie.fr
%LISTIC, Polytech'Savoie-Annecy, Universit\'e de Savoie
%LaTeX2e
%encoding : utf8

\section{Control of DDIFS}\label{sec:control}
%pmx2009-04-16: placer cette figure en amont pour qu'elle soit en t\^ete de la page 2!
% % % % % \begin{figure*}
% % % % % \begin{center}
% % % % % \includegraphics[scale=0.3]{fig/twolevelscontrol.eps}
% % % % % \end{center}
% % % % % \caption{Information Fusion Process (IFP)- Information Fusion System (IFS) relationship\label{fig:twolevelscontrol}}
% % % % % \end{figure*}

%pmxbegin2009-04-16: cette figure doit  \^etre plac\'ee dans la section reconfiguration
%finalement elle n'est pas ins\'er\'ee dans le document
% % % % olivier 06-04-09 : addition of fault tree
% % % \begin{figure}
% % % \begin{center}
% % % \includegraphics[scale=0.3]{fig/fault.tree.eps}
% % % \end{center}
% % % \caption{Execution and control sub-systems fault tree:
% % % each box represents a fault.
% % % If the fault is not caught it induces an error and
% % % then a fault to higher levels.\label{fig:faulttree}}
% % % \end{figure}
%pmxend2009-04-16

Figure~\ref{fig:twolevelscontrol} presents the two levels view
architecture of our proposal:
\begin{itemize}
\item
the fusion graph, i.e. the fusion nodes (or fusion functions) and
the connections between their input and output ports;
\item
the assignment of the fusion functions to the elements of the
fusion run-time system.
\end{itemize}
The run-time DDIFS is controlled for what concerns
error recovery and quality improvement.

Our system is developed in such a way that it checks
for the IFS consistency i.e.
the availability of the communication network between runtime sub-systems
and the availability of these sub-systems.
To this end, software sensors installed into the system provide
both quantitative and qualitative measurements.
The time interval between two executions of a fusion function
or the amount of data present in some point of the system are such measurements.
Thanks to these software sensors, errors and failures are detected and the system
updates itself by changing its configuration in order to correct them.

The quality improvement is based on a
Generalized Stochastic Petri Net (GSPN)~[\cite{IBAMBoDo98}] model of the configuration.
We build a GSPN of the run-time system
from which we derive a set of performance/dependability steady-state rewards $(r_n)_{1\leq n\leq N}$.
These rewards (CPU utilisation, response times, etc.) are computed
from the Markov chain underlying the GSPN,
with a tool, like GreatSPN~[\cite{ARChFrGaRi95}]
running on one of the host systems of the IFS.
Details of the performance analysis of our system will be presented in a future paper.
In short, we build a total order between configurations based on the rewards $(r_n)$.

% % % The set of configurations $c$ equipped with their reward vectors
% % % $R(c)=(r_{1}(c), \ldots, r_{N}(c))$ is at least a partially ordered set:
% % % $c \preceq c'$ iff $R(c) \leq R(c')$ where $\leq$ is the component-wise vector
% % % comparison.
% % % In the present work, we have implemented a generic total order
% % % between configurations defined through a function $\delta(R,R')$:
% % % $c \preceq c'$ iff $\delta(R(c),R'(c)) \geq 0$.
% % % $\delta$ may be redefined at run-time, and its default value is
% % % $\delta(R(c),R'(c)) =\sum_{n=1}^N \alpha_n (r_{n} - r'_{n})$,
% % % with $(\alpha_i)_{i=1,\cdots,n}$ a sequence of positive weights.

% end of file control.tex

%file: implementation.tex
%description: control of distributed information fusion system, dcds09
%date: 2009-04-16
%author: Olivier PASSALACQUA
%email: olivier.passalacqua@univ-savoie.fr
%LISTIC, Polytech'Savoie-Annecy, Universit\'e de Savoie
%LaTeX2e
%encoding : utf8

\section{Implementation}\label{sec:implementation}
%%%%\begin{figure}
%%%%\begin{center}
%%%%\includegraphics[scale=0.5]{fig/fusion.node.terms.eps}
%%%%\end{center}
%%%%\caption{Implementation details - Runtime frameworks hierarchy.
%%%%\label{fig:implementation.hierarchy}}
%%%%\end{figure}

%%%%%%%%%%%\begin{figure}
%%%%%%%%%%%\begin{center}
%%%%%%%%%%%\includegraphics[scale=0.4]{fig/fusion.node.container.eps}
%%%%%%%%%%%\end{center}
%%%%%%%%%%%\caption{Implementation details - Fusion node implementation.
%%%%%%%%%%%\label{fig:implementation.fusionnode}}
%%%%%%%%%%%\end{figure}

\begin{figure}
\begin{center}
\includegraphics[scale=0.6]{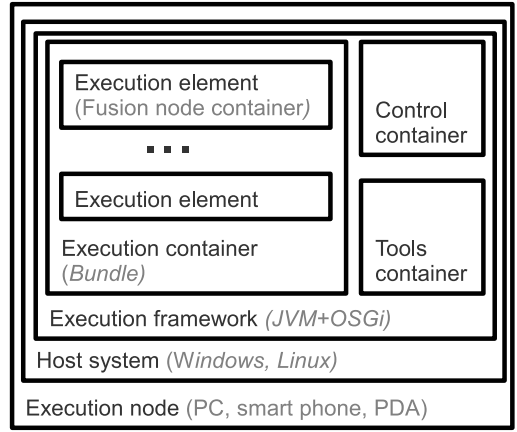}
\end{center}
\caption{Implementation details - Runtime frameworks hierarchy.
\label{fig:implementation.hierarchy}}
\end{figure}
To take into account modern architectures,
our model distinguishes four hierarchical levels
in an IFS (Figure~\ref{fig:implementation.hierarchy}).
The lowest level is the physical machine level such as a Personal Computer,
a smart-phone, etc.
Each machine supports one or several host operating systems (as in virtualised systems).
On top of a given host, an EF defines
the fundamental architectural element of our IFS,
assuming that every EF owns a unique (IP address, port number) pair.
Each EF hosts the two sub-systems of our solution:
an execution sub-system and a control sub-system.

The intend of our system is to take advantage of the skills
of both information fusion experts and developers.
Thus while the (information fusion) designer defines
the data-flow graph that represents the fusion process
through a graphical interface,
developers may implement the execution codes of the fusion functions.
In this way, the designer expresses specifications on the fusion process,
and the developer only writes the selected fusion method
Both  do not take care about the deployment
nor the modifications of the run-time system.

An implementation of the FP, also termed as a \emph{configuration},
is defined by the choice of all the implementation elements translating the FP:
fusion node implementations (see below),
assignment of the fusion nodes to EF (a fusion node is assigned to one EF),
ports links mapping between fusion nodes.
It assumes that EFs are linked through an undirected connected  IP network
(the \emph{execution graph})
and that all the connections between fusion node ports are carried by this network ($N$).
If the output $y_k$ of a function $f$ assigned to the EF $e$
is the input $x_i$ of $f'$ assigned to $e'$,
then the configuration defines the path between $e$ and $e'$ in $N$.
This path may use intermediate EF only used to connect the source and the
destination EFs.
%pmx2009-04-16
% % When the fusion process has been described and the fusion functions
% % have been developed, the control system uses the data-flow graph
% % and send requests to the EFs to deploy the fusion nodes.
As soon as a configuration is defined, it is deployed by the control sub-systems
of the EFs.

An execution sub-system manages the execution elements corresponding to fusion nodes.
In contrast, a control sub-system manages the execution sub-system:
monitoring and analysis of the FP execution and reconfigurations of the IFS.

\subsection{Fusion Node implementation}
At the fusion process level, a fusion node consists of three sets of ports
(inputs, parameters, outputs) and a mathematical description of the fusion function.
Our model assumes that at least one implementation of each fusion function is available
in the system and that all the implementations are valid
i.e. they conform to their mathematical specification.
\begin{figure}
\begin{center}
\includegraphics[scale=0.7]{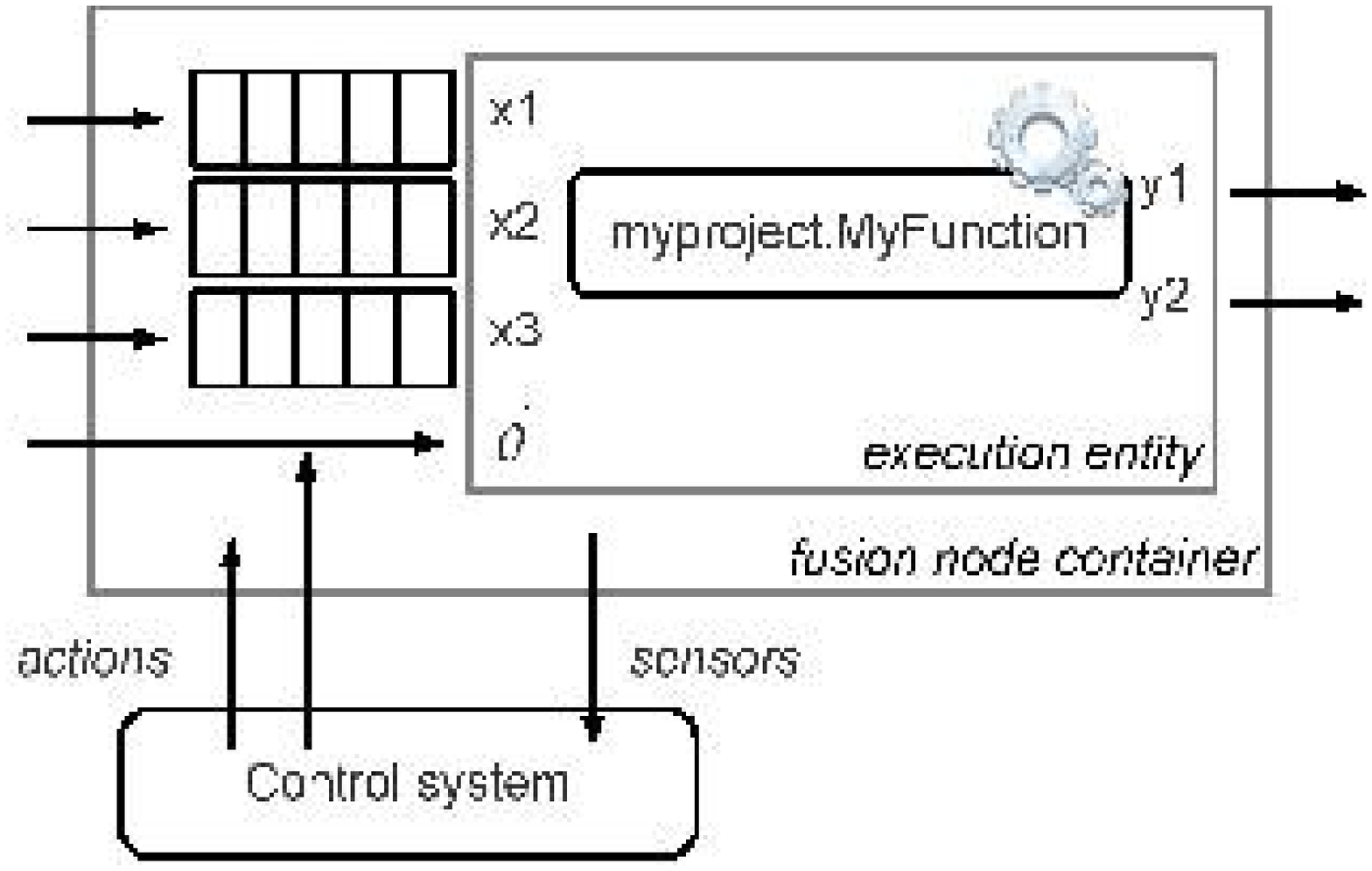}
\end{center}
\caption{Implementation details - Fusion node implementation.
\label{fig:implementation.fusionnode}}
\end{figure}
To manage the execution of a fusion node related tasks,
we introduce a fusion node container (Figure~\ref{fig:implementation.fusionnode}).
It controls the implementation of the fusion function, termed the execution
entity; it updates the value of the control parameters,
% % % olivier 06-04-09 : sentence update
% % and it transfers the software sensors outputs to other input ports.
and it transfers the software sensors values to the rest of the control system.

\subsection{Execution mechanism}
Implementation of the input port discrete semantics is based on queues
storing discrete data,
one queue being bound to each input port of a fusion node.
Each computation of the fusion function un-queues one data item of each queue
and uses the latched values of the parameters.
Moreover, according to the \emph{Best effort} policy defined in Section~\ref{sec:transition}
the system tries not to lose data when unreachable %destination
fusion nodes are detected.
Hence, the execution of the fusion function is started
iff the following conditions are satisfied:
\begin{itemize}
\item no input queue is empty;
\item the result of the computation may be sent to all its consumers.
\end{itemize}

The first condition is required to provide a value of each input port to the
execution entity.
The second condition is a design decision: we do not start a computation
of the fusion function if we are not sure (to the best of our knowledge!)
that the results could be processed by their receivers.

% % olivier 02-04-09 : section moved from example section after DCDS review
\subsection{Implementation details}
An implementation of our project is already deployed
in order to experiment our results and
was used to run a video conference smart-room[\cite{IPWaToSaPaBeHuMo08}].

The current version of our system is deployed over
OSGi~[\cite{OSGI4}] platforms~[\cite{IPReAlRo07}]
linked by standard networks (LAN, Wi-Fi, Bluetooth).
EFs are implemented as OSGi platforms and
are linked together by R-OSGi services.
The sub-systems are presently implemented as bundles,
respectively for the control and the execution sub-systems,
installed and started on each OSGi framework.
Thanks to this structure the control sub-system can easily manage
the execution sub-system especially during transitions.

% % % % olivier 06-04-09 : moved to the begin of the section
% % The intend of our system is to take advantage of the experience
% % of both information fusion experts and developers.
% % Thus while the (information fusion) designer defines
% % the data-flow graph that represents the fusion process
% % through a graphical interface,
% % developers may implement the execution codes of the fusion functions.
% % In this way, the designer expresses specifications on the fusion process,
% % and the developer only writes the selected fusion method and does not take
% % care about deployment nor transition mechanisms.

% % When the fusion process has been described and the fusion functions
% % have been developed, the control system uses the data-flow graph
% % and send requests to the EFs to deploy the fusion nodes.

The execution entities (code of the fusion nodes) defined by the configuration
and the control elements are then registered as services in the OSGi framework
and act together thought services discovery and requests.

%end of file implementation.tex

%file: reconfiguration.tex
%encoding : utf8
%description: control of distributed information fusion system, dcds09
%date: 2009-04-16
%author: Olivier PASSALACQUA
%email: olivier.passalacqua@univ-savoie.fr
%LISTIC, Polytech'Savoie-Annecy, Universit\'e de Savoie
%LaTeX2e

\section{Reconfiguration of DDIFS}\label{sec:reconfiguration}

The reconfiguration process, that is to say, the update of the implementation of
the IFP on the IFS is made up of three phases:
\begin{itemize}
  \item setting of the reconfiguration strategy,
  \item selecting a new configuration,
  \item deploying and starting the new configuration.
\end{itemize}

\subsection{Reconfiguration setup}
The reconfiguration of the system is twofold:
correction of a configuration in order to restore the fusion process after
an execution failure or an execution error,
or improvement of the current configuration.
The possible modification of a configuration relates to
fusion function implementations, function assignments to EF
or to input-output ports link mappings.

\subsubsection{Configuration errors and failures}
% % % \begin{figure}
% % % \begin{center}
% % % \includegraphics[scale=0.3]{fig/fault.tree.eps}
% % % \end{center}
% % % \caption{Execution and control sub-systems fault tree\label{fig:faulttree}}
% % % \end{figure}
The system is said to be in a failure state [\cite{ARAvLaRaLa04}]
when one of its outputs cannot produce a result.
Such a failure derives from an error.
% % % Figure~\ref{fig:faulttree} gives the fault tree of the three possible
% % % faults of a configuration.
% % % Each box represents a fault at a given level leading,
% % % when not caught, to an error and then to a fault to the higher level.
Our system tries to avoid as much as possible system failures,
by detecting errors before they lead to failures.
We detect two kinds of error:
\begin{itemize}
\item
Inter-execution framework errors:
an EF disappears or a communication channel is broken;
\item
Intra-execution framework errors:
%%%it assumes that the control system is able to distinguish
errors due to fusion node interactions (i.e. deadlock)
and internal fusion node errors (for instance arithmetic overflow, time-out).
\end{itemize}
%pmxbegin2009-04-16

As mentioned in Section~\ref{sec:control}, software sensors, mostly throwing
a Java exception, are used to detect these errors.
Inter-execution framework errors are detected through monitoring of the communication
links with programmed acknowledgments.
Intra-execution framework errors throw Java exceptions caught by the control system
of the execution framework.
%pmxend2009-04-16

\subsubsection{Configuration improvement}
Even if there is no error,
the control system tries to improve the runtime system permanently.
To this end, firstly the control system periodically senses the model parameters
from the execution system and computes the rewards associated to the model.
If a significant variation is stated between two or more computations,
the control system triggers a search for a new, and hopefully ``better", configuration.
Secondly, the control system also launches a new configuration search when it
detects a variation in the
%pmxbegin2009-04-16
available
%pmxend2009-04-16
runtime environment, for example due to a new
available EF.

\subsection{Selection of a new configuration}
As soon as the need for a reconfiguration is launched, the system has to search for a
new configuration.
The search algorithm takes the description of the IFP and
the constraints between the fusion nodes and the EFs as inputs.
In fact, the selection of a new configuration due to reconfiguration
involves the same steps as the initial configuration selection.

The search is based on a Constraint Programming approach
already proposed by several researchers~[\cite{ARZhSaBeWaSi08,IPAnWoHaKa07}]
in the context of the application component placement problem~[\cite{IPKiIvKa03}].
The IFP provides a set of constraints such as the set of fusion nodes,
the required links between output and input fusion nodes.
The IFS constraints the possible configurations in several ways:
\begin{itemize}
\item
each fusion function can be deployed on a subset only of the EF;
\item
connections between EF are fixed and given through an execution graph.
We call "channel" a direct connection between two EF $e$ and $e'$.
\item
for a given link $l$ between $y_k$ and $x'_i$, we must select a
chain $((e_m, e_{m+1}))_{1\leq m \leq M}$ of channels in the execution graph
such that $e_1=e$ and $e_{M+1}=e'$.
\item
a given EF must have enough memory resources to be able to run the fusion functions
assigned to it.
\end{itemize}
Hence, for a given assignment of fusion nodes to EF together with the paths in the
execution graph derived from links between fusion nodes,
we are able to express automatically  a set of constraints on these assignments.

There are in general several paths in the execution graph.
For each path, we can define a "cost" based for instance on the number of its channels
, i.e. its "length", an effective usage cost, etc.
We assume that all links from $f$ to $f'$ use the same path in the execution graph.
Although we have designed our search in  a generic cost way,
in the present work we have implemented only the "length cost".
We then fix the "best" path as the shortest path (in the cost meaning)
in the execution graph between the EF of the linked fusion nodes.
These shortest paths between EF are pre-computed for a given IFS
with the Floyd-Roy-Warshall algorithm (see for instance~[\cite{BOCoLeRiSt01}]
and they are used by the Constraint Satisfaction Problem (CSP) solver.
The CSP is now well defined and its variables are the assignments of fusion functions
to the EF.

We can also add an optimisation criterium to the previous CSP to
take into account the two antagonistic properties of a configuration:
\begin{itemize}
\item
global maximal usage of the set of EFs: deploy the fusion nodes on as much as possible
EF;
\item
global ``short" physical communication between output and input fusion functions:
deploy linked fusion functions on "neighbouring" EFs.
\end{itemize}
To do so, we define a generic cost function $C$ of a configuration:
$C = h(C_d, C_c)$ where $C_d$ is a cost associated to the assignment of
the fusion functions,
and $C_c$ is a cost associated to the mapping of
the links to the channel chains and $h$ a composition function.
Note that $C_d$ should be increasing with the ``density" of the
fusion functions on the EF.
For instance, $C_d$ could be:
$$
C_d = \max_{e \in E}\{n(e)\} - \min_{e \in E}\{n(e)\}
$$
with $n(e)$ being the number of fusion functions assigned to the node $e$
and $E$ the set of EFs.
For $C_c$ we can take a weighed ($\alpha_u$) sum of the number of used channels:
$$
C_c = \sum_{u \in U} \alpha_u n(u)
$$
where $n(u)$ is the number of links (between fusion function ports) using the
channel $u$ and $U$ is the set of channels of the execution graph.
$C_c$ should also be increasing with the ``distribution" of the fusion functions
in the IFS.

%%Whatever the reason for reconfiguration, the search algorithm
%%could be either a centralised or a distributed algorithm.
%%Work is in progress to compare both kinds of solutions
%%and will be presented in the final version of the paper.

Finally, we send the problem to a CSP - with optimization- solver to get the configuration.
We used the Choco solver which is available at \url{http://www.emn.fr/x-info/choco-solver/doku.php}.
For the moment, the solver is run on one of several EF defined in a configuration file
of the system.

\subsubsection{Error recovery}
In the case of an error recovery, the control system selects as soon as
possible a configuration compatible with the available resources.
Hence, the search algorithm selects the first admissible
configuration. Searching for a better configuration is performed during
the next configuration improvement step.

\subsubsection{Configuration improvement}
In this case, the control system searches for another ``significantly better"
configuration.
To this end, from the current configuration, it derives a model
of the configuration and computes its rewards as explained in Section~\ref{sec:control}.
Then it throws a search for a new configuration as a CSP with an optimisation
based on comparison of the rewards.

\subsection{Transition between configurations}\label{sec:transition}
Independently from the reconfiguration strategy, the system applies the new
configuration $\name{NC}$ from the current configuration $\name{CC}$ as follows.
The system updates the location of the fusion node implementations,
in case of new function assignments, and/or updates the execution entities, in
case of implementation swaps.
In both cases the system behaves according to a best effort policy.

\subsubsection{Best effort}
The system tries to prevent loss of data-flow driven
information present in it and already partially processed.
To do so it is assumed that two different implementations of
the same fusion function have the same semantics in the IFP.
Thus, input data of a fusion node may come from computations
ran in any configuration without invalidating the next produced results.

Let $e'$ (in $\name{NC}$) be an updated version of the EF $e$ (in $\name{CC}$).
If $e'$ can restore some data from the state of $e$,
for instance by swapping an implementation of a fusion function,
data waiting in the input queues are processed by the new execution entities.
Hence there is no loss of data in this case.

On the contrary, partially processed data present on $e$ are lost
in case of assignments of the fusion functions of $e$ to an EF different from $e'$.
In such a reconfiguration, the links between the functions are mapped into new paths according to
the new location of the fusion node containers.
Each fusion node container, and therefore its data, is destroyed on $e$
and instantiated on new EFs with empty input queues.

The current transition mechanism can be extended in order to
cope special properties such as synchronization.
In such a case and only when partially processed data are lost during
the transition, some remaining data present in other EFs
may be out of synchronization.
To deal with this property, two extensions have to be added to our model:
a label linked to the data that identifies the synchronization criterium,
i.e. the time of sensors read, and a control element that deletes a data
which is out of synchronization after a transition.
Synchronization mechanism is presently not implemented in our system
but is already defined in our model.

% end of file reconfiguration.tex

%file: example.tex
%encoding : utf8
%description: control and verification of distributed information fusion system, dcds09
%date: 2009-01-25
%author: Olivier PASSALACQUA
%email: olivier.passalacqua@univ-savoie.fr
%LISTIC, Polytech'Savoie-Annecy, Universit\'e de Savoie
%LaTeX2e

% % % olivier : 02-04-09 : correction after DCDS review, only one example
\section{Application example}\label{sec:examples}

\begin{figure}
\begin{center}
\includegraphics[scale=0.8]{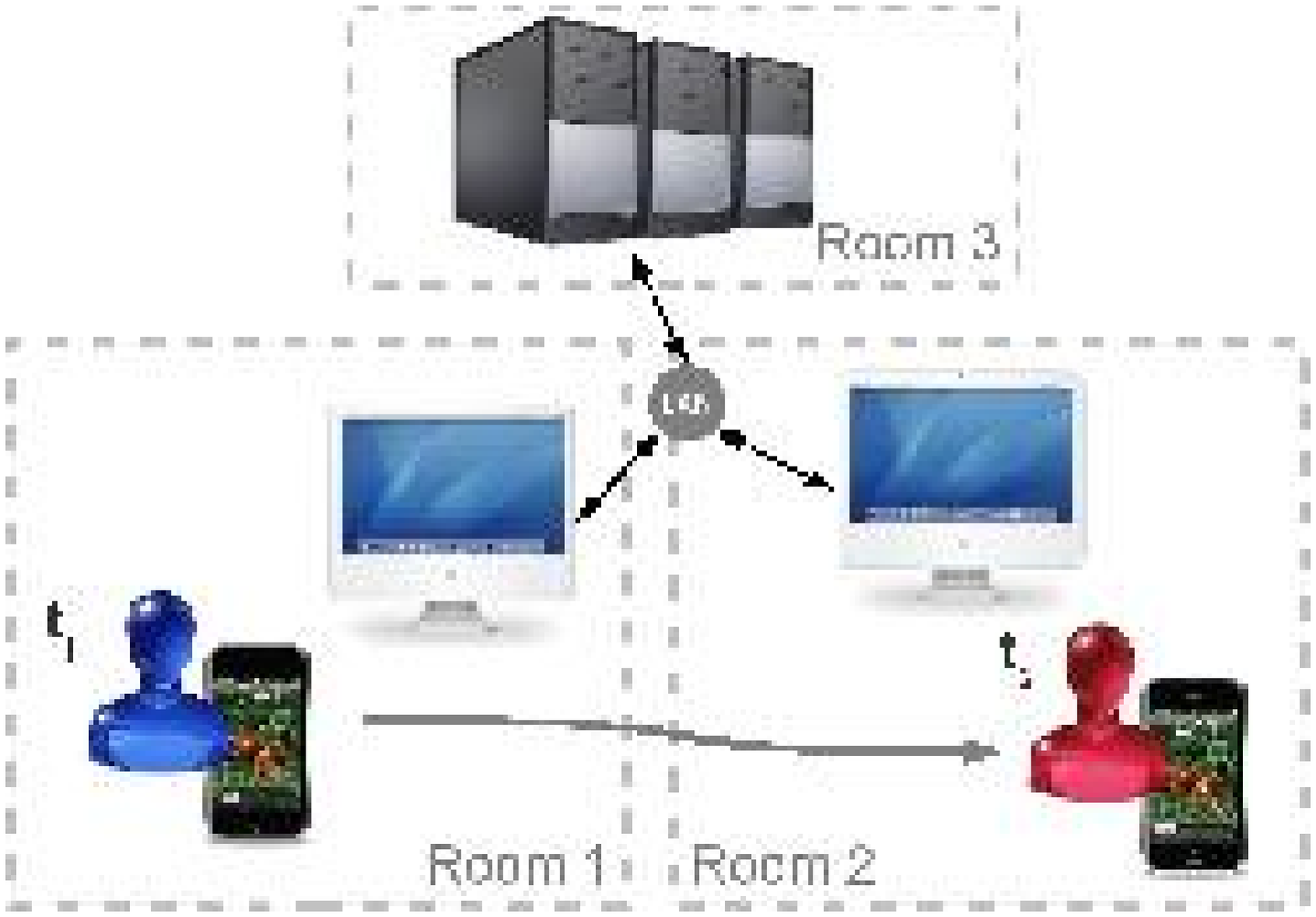}
\end{center}
\caption{Passer-by example of reconfiguration - architecture view.
% The user moves from room1 to room2 that induces the loss of network
% connection between his smart phone and the system.
}
\label{fig:example.passerby.archi}
\end{figure}

\begin{figure}
\begin{center}
\includegraphics[scale=0.4]{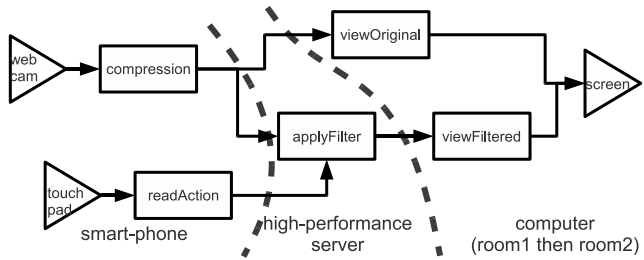}
\end{center}
\caption{Passer-by example of reconfiguration - process view.
% In order to recover the execution of the fusion process the
% smart phone is connected into room2 and the fusion function
% are assigned to the second computer.
}
\label{fig:example.passerby.proc}
\end{figure}

% % % % % olivier 02-04-09 : figure removed after DCDS review
% % % \begin{figure}
% % % \begin{center}
% % % \includegraphics[scale=0.6]{fig/example2.eps}
% % % \end{center}
% % % \caption{Implementation swap example of reconfiguration.}
% % % \label{fig:example.swap}
% % % \end{figure}

\subsection{Passer-by example}
In the passer-by example (Figures~\ref{fig:example.passerby.archi} and \ref{fig:example.passerby.proc})
the computers respectively located in room1 and room2
%%front of the user
display two videos.
The first one is the original view of the user's head (taken from a smart-phone).%web-cam).
The second one is a filtered view of the original view,
and the selected filter is chosen by the user through a touch-pad.
In the first configuration ($t_1$) the user is in the room1 and
his smart-phone produces a video of his head.
The video is processed by powerful computers located in room3 and
the two videos are both displayed on the screen in room1.
While the user is moving between the two rooms, the system detects that the
connection with the smart-phone is lost and tries to reconfigure itself.
As soon as the communication is restored with the smart-phone,
the second configuration ($t_2$) executes the same fusion process but the last
functions are assigned to the second computer in room2.

\section{Conclusion}\label{sec:conclu}
In this paper, we have presented problems raised by reconfiguration features
of a runtime model for controlled dynamic distributed information fusion system (CDDIFS).
Dynamicity comes from the changing runtime support of our systems.
Reconfiguring a CDDIFS system means either correcting it because of an error
or a failure, or else increasing its quality of service.
Our proposal is based on several functional components:
- monitoring of the networked run-time system providing indication on the availability of
the sub-systems and raw performance measures;
- computation of a dependability model of the configurations running the
Distributed Information Fusion System.
- definition of a Constraint Satisfaction Problem (CSP) modeling the placement of the
fusion functions onto the Execution Framework;
We take advantage of efficient solvers for both
computation of performance indices (rewards) based on the dependability model
and resolution of the CSP.
We are hence able to reconfigure in the ``best way'' with respect to a given
set of criteria, our CDDIFS.
Future work will deal first with full automation of all the components of our
framework and installation of the system on top of other middleware systems
like networked J2EE servers.
We are also testing our framework with several kinds of Information Fusion Processes
such as scientific computation systems, energy control systems, mechatronic systems.

% end of file conclusion.tex

%\bibliographystyle{ifacconf}
\bibliographystyle{alpha}
\bibliography{biblio}

\end{document}